\documentclass[reprint,superscriptaddress,aps,twocolumn,showpacs]{revtex4-1}
\usepackage{amssymb}
\usepackage{amsmath}
\usepackage{graphicx}
\usepackage{multirow}
\usepackage{hyperref}
\usepackage{xcolor}
\usepackage{tabularx}
\allowdisplaybreaks

\begin{document}
\title{Systematic study of exotic $1^{-+}$ tetraquark spectroscopy}

\author{Kai Xu}
\email[]{gxukai1123@gmail.com}
\author{Zheng Zhao}
\email[]{zhaozheng1022@hotmail.com}
\author{Nattapat Tagsinsit}
\email[]{nattapattagsinsit@gmail.com}
\author{Attaphon Kaewsnod}
\affiliation{School of Physics and Center of Excellence in High Energy Physics and Astrophysics, Suranaree University of Technology, Nakhon Ratchasima 30000, Thailand}
\author{Ayut Limphirat}
\author{Christoph Herold}
\email[]{herold@g.sut.ac.th}
\author{Yupeng Yan}
\email[]{yupeng@sut.ac.th}
\affiliation{School of Physics and Center of Excellence in High Energy Physics and Astrophysics, Suranaree University of Technology, Nakhon Ratchasima 30000, Thailand}

\date{\today}

\begin{abstract}
The masses of exotic quantum-number $1^{-+}$ compact tetraquark states are calculated in a constituent quark model, where a Cornell-like potential is employed as the central potential, spin-spin and spin-orbit coupling derived from the Breit-Fermi interaction are treated as hyperfine corrections, and model parameters are taken from previous works. The ground state $1^{-+}$ P-wave tetraquarks are predicted at 1.9, 4.2, and 6.6~GeV for the light, charmoniumlike, and fully charm sectors, respectively. 
The decay width ratios of $1^{-+}$ tetraquark states are calculated for two-body strong decay channels within the rearrangement mechanism, including $\omega h_1$ and $\eta f_1$ for isospin $I=0$ light tetraquarks, $\rho h_1$ and $\pi f_1$ for isospin $I=1$ light tetraquarks, $\pi/\eta+\chi_{c1}$ and $\rho/\omega + h_c$ for charmoniumlike tetraquarks, and $\eta_c \chi_{c1}$ and $J/\psi h_c$ for fully charm tetraquarks. The theoretical results are compared with the observed exotic $1^{-+}$ states, and promising search channels for $1^{-+}$ tetraquarks are discussed.
The work suggests that $\eta_1(1855)$ is unlikely to be a compact tetraquark state.

\end{abstract}

\maketitle

\section{Introduction}\label{sec:Int}
Understanding the spectrum and internal structure of hadrons remains a central goal of quantum chromodynamics (QCD), and states beyond the conventional $q\bar q$ mesons and $qqq$ baryons, such as multiquark configurations and hybrids, are now an active frontier of experiment and theory.

charmoniumlike states with total angular momentum $J=1$ have been widely searched over the last two decades and are promising P-wave tetraquark candidates. Several exotic states in the charmonium region, known as Y states with $1^{--}$ quantum numbers that mix with conventional $\psi$ mesons, could be P-wave tetraquark candidates: Y(4230)~\cite{BaBar:2005hhc}, Y(4360)~\cite{Belle:2007umv}, Y(4660)~\cite{BaBar:2012hpr}, the more recent Y(4484)~\cite{BESIII:2022joj}, and Y(4544)~\cite{BESIII:2024jzg}.

In conventional mesons composed of a quark and an antiquark $q\bar{q}$, the quantum numbers $J^{PC}$ follow simple rules: parity $P=(-1)^{L+1}$ and charge conjugation $C=(-1)^{L+S}$, where L is orbital angular momentum and S is total spin~\cite{PDG}. These rules allow common states such as $1^{--}$ (the $\rho$ meson), $1^{+-}$ (the $h_1$ meson), or $1^{++}$ (the $f_1$ meson) but not $1^{-+}$. Thus, a $1^{-+}$ state, once observed, is naturally considered to have a nonconventional structure, such as a hybrid meson, hadronic molecule, or tetraquark.

The experimental status of exotic $1^{-+}$ states has advanced through complementary approaches. The existing data provide a foundation for evaluating their properties, which we briefly review below:

The vertex spectrometer (VES) experiment observed the exotic $1^{-+}$ $\pi_1(1400)$ state in $\pi^- p \to \pi \eta n$ decays at 100 GeV~\cite{IHEP-Brussels-LosAlamos-AnnecyLAPP:1988iqi}, with a mass of $1406 \pm 20$ MeV and a width of $180 \pm 30$ MeV. The analysis of the $\eta\pi$ channel showed a clear exotic signal inconsistent with conventional $q\bar{q}$ states.

The $\pi(1600)$ state was first observed by the Brookhaven E852 experiment~\cite{E852:1998mbq}, which identified an exotic $1^{-+}$ resonance in the $\rho\pi$ channel. The analysis found a mass of $1593 \pm 8^{+29}_{-47}$ MeV and a width of $168 \pm 20^{+150}_{-12}$ MeV. Later, E852~\cite{E852:2004gpn} observed two exotic $1^{-+}$ resonances in $\pi^- p \to f_1(1285)\pi^- p$ collisions at 18~GeV: the known $\pi_1(1600)$ with mass $M = 1709 \pm 24 \pm 41$ MeV and width $\Gamma = 403 \pm 80 \pm 115$ MeV, and a new $\pi_1(2015)$ state at $M = 2001 \pm 30 \pm 92$ MeV with width $\Gamma = 333 \pm 52 \pm 49$ MeV.

In the updated Particle Data Group (PDG) review, $\pi(1400)$ and $\pi(1600)$ are suggested to be a single state, as coupled-channel analyses favor one broad isovector resonance in the 1.4--1.6~GeV region~\cite{PDG}. Of the reported candidates, the $\pi_1(2015)$ has only been observed by the E852 experiment~\cite{E852:2004gpn}, and its existence remains inconclusive.

The BESIII Collaboration recently observed the isoscalar exotic state $\eta_1(1855)$ with $1^{-+}$ in $J/\psi \to \gamma \eta \eta'$ decays~\cite{BESIII:2022riz}, measuring its mass as $(1855 \pm 9^{+6}_{-1})$ MeV, its width as $(188 \pm 18^{+3}_{-8})$ MeV with $>19\sigma$ significance, and the product branching fraction $\mathcal{B}(J/\psi \to \gamma \eta_1(1855)) \times \mathcal{B}(\eta_1(1855) \to \eta \eta') = (2.70 \pm 0.41^{+0.16}_{-0.35}) \times 10^{-6}$. The narrow width of $\eta_1(1855)$ contrasts sharply with the broad $\pi_1(1600)$, suggesting different underlying dynamics between these exotic states.

In this work, we study the mass spectrum and decay properties of the $1^{-+}$ P-wave light, charmoniumlike, and fully charm compact tetraquark states within a constituent quark model framework.

The paper is organized as follows. In Sec.~\ref{sec:TM}, all possible configurations of the color, spin, flavor, and spatial degrees of freedom of tetraquark states are introduced, and a constituent quark model developed in our previous work~\cite{Zhao:2020jvl,Zhao:2021jss,Zhao:2025kno} is briefly reviewed. In Sec.~\ref{sec:RND}, the $1^{-+}$ P-wave tetraquark mass spectra and relative decay ratios are evaluated within the constituent quark model. The theoretical results are compared with experimental data, and discussions of $\pi_1(2015)$ and $\eta_1(1855)$ are provided. A summary is given in Sec.~\ref{sec:SUM}.

\section{\label{sec:TM}THEORETICAL MODEL}

\subsection{\label{sec:TMA}Quark configurations of tetraquark}

The construction of tetraquark states follows the principle that a tetraquark must be a color singlet, meaning the color wave function must correspond to a $[222]_1$ singlet within the $SU_c(3)$ group. Furthermore, the total wave function of any cluster of identical fermions must be antisymmetric.

In this work, we consider $1^{-+}$ light ($q_1q_2\bar q_3 \bar q_4$), charmoniumlike ($q_1c_2\bar q_3 \bar c_4$), and fully charm ($c_1c_2\bar c_3 \bar c_4$) tetraquark states, where $q$ and $c$ denote a light quark and a charm quark, and $\bar q$ and $\bar c$ denote a light antiquark and a charm antiquark, respectively.

We employ a diquark-antidiquark coupling scheme and classify each cluster by SU(3) color Young tableaux. The permutation symmetry of the two-quark cluster ($qq$, $qc$, or $cc$) in tetraquark states is described by the Young tableaux $[2]_6$ and $[11]_{\bar 3}$ of the $SU_c(3)$ group, while the color configuration of the two-antiquark cluster ($\bar q\bar q$, $\bar q\bar c$, or $\bar c\bar c$) consists of a $[211]_3$ triplet and a $[22]_{\bar 6}$ antisextet.

Thus, a $[222]_1$ color singlet for tetraquark states requires the following color configurations: $[2]_6\otimes[22]_{\bar 6}$ and $[11]_{\bar 3}\otimes[211]_3$, which correspond to the color sextet-antisextet ($6_c \otimes \bar 6_c$) and triplet-antitriplet ($\bar 3_c \otimes 3_c$) configurations, respectively.

The $1^{-+}$ tetraquark states correspond to $L=1$, $S=1$ configurations according to the parity ($P=(-1)^L$) and charge-conjugation ($C=(-1)^{L+S}$) quantum numbers, where $L$ is the total orbital angular momentum and $S$ is the total spin of the tetraquark. Thus, the $1^{-+}$ tetraquark requires an odd $L$, with the lowest possible value being $L=1$. The complete basis is constructed by coupling harmonic-oscillator wave functions, which takes the general form:

\begin{eqnarray}\label{eqn::spatial}
\psi_{NL} &=
\sum_{{n_{\chi_i},l_{\chi_i}}}
 A(n_{\chi_1},n_{\chi_2},n_{\chi_3},l_{\chi_1},l_{\chi_2},l_{\chi_3}) \nonumber \\
& \times \psi_{n_{\chi_1}l_{\chi_1}}(\vec \chi_1\,) \otimes\psi_{n_{\chi_2}l_{\chi_2}}(\vec \chi_2\,)\otimes\psi_{n_{\chi_3}l_{\chi_3}}(\vec \chi_3)
\end{eqnarray}
where $\psi_{n_{\chi_i}l_{\chi_i}}$ are harmonic oscillator wave functions. The sum ${n_{\chi_i},l_{\chi_i}}$ runs over $n_{\chi_1},n_{\chi_2},n_{\chi_3}, l_{\chi_1},l_{\chi_2},l_{\chi_3}$. $N$ and $L$ are, respectively, the total principal quantum number and total angular momentum, where $N= (2n_{\chi_1}+ l_{\chi_1})+(2n_{\chi_2}+ l_{\chi_2})+(2n_{\chi_3}+l_{\chi_3})$, and $L=l_{\chi_1}+l_{\chi_2}+l_{\chi_3}=1$ for the P wave tetraquark. The relative Jacobi coordinates $\vec \chi_1$, $\vec \chi_2$, and $\vec \chi_3$ are defined as:
\begin{flalign}\label{eqn::jacobi}
&\vec \chi_1=\frac{1}{\sqrt 2}(\vec r_1-\vec r_2), \nonumber \\
&\vec \chi_2=\frac{1}{\sqrt 2}(\vec r_3-\vec r_4), \nonumber \\
&\vec \chi_3=\frac{m_1\vec r_1+m_2\vec r_2}{m_1+m_2}-\frac{m_3\vec r_3+m_4\vec r_4}{m_3+m_4}, 
\end{flalign}
where $\vec{r}_{j}$ and $m_j$ are the coordinate and the mass of the jth quark, respectively. The reduced quark masses are defined as:
\begin{flalign}\label{eqn::rqm}
&u_1=\frac{2m_1m_2}{m_1+m_2}, \nonumber \\
&u_2=\frac{2m_3m_4}{m_3+m_4}, \nonumber \\
&u_3=\frac{(m_1+m_2)(m_3+m_4)}{m_1+m_2+m_3+m_4},
\end{flalign}
The size of complete bases can be expanded by increasing the principal quantum number $N$. The complete bases with $N=1$ are listed in Table~\ref{spatialbasis} for reference.  In this work, we employ complete bases of $N\leq13$ to calculate the mass spectrum.
\begin{table}[htbp]
\caption{\label{spatialbasis} The complete bases of tetraquark with the lowest quantum number, $N=2n+L=1$ and $L=1$.}
\begin{ruledtabular}
\begin{tabular}{ccc} 
\multirow{2}{*}{$\Psi_{NL}$} & Bases & \multirow{2}{*}{$\psi_{n_{\chi_1}l_{\chi_1}}(\vec\chi_1\,)\psi_{n_{\chi_2}l_{\chi_2}}(\vec\chi_2\,)\psi_{n_{\chi_3}l_{\chi_3}}(\vec\chi_3\,)$} \\
&  \{$l_{\chi_1},l_{\chi_2},l_{\chi_3}$\}  & \\
\noalign{\vskip 5pt}
\multirow{3}{*}{$\Psi_{11}$} & \{0,\,0,\,1\} & $\psi_{0,0}(\vec\chi_1\,)\psi_{0,0}(\vec\chi_2\,)\psi_{0,1}(\vec\chi_3\,)$,\\
& \{0,\,1,\,0\}  &  $\psi_{0,0}(\vec\chi_1\,)\psi_{0,1}(\vec\chi_2\,)\psi_{0,0}(\vec\chi_3\,)$, \\
& \{1,\,0,\,0\}  & $\psi_{0,1}(\vec\chi_1\,)\psi_{0,0}(\vec\chi_2\,)\psi_{0,0}(\vec\chi_3\,)$ \\ 
\end{tabular}
\end{ruledtabular}
\end{table}

For $S=1$ configurations, the possible spin combinations are $ \left[\psi_{[s=1]}^{q_1q_2}\otimes\psi_{[s=1]}^{\bar q_3 \bar q_4}\right]_{S=1}$, $\psi_{[s=1]}^{q_1q_2}\otimes\psi_{[s=0]}^{\bar q_3 \bar q_4}$, and $\psi_{[s=0]}^{q_1q_2}\otimes\psi_{[s=1]}^{\bar q_3 \bar q_4}$.
As a $qq\bar{q}\bar{q}$ or $cc\bar{c}\bar{c}$ tetraquark must be a color singlet and simultaneously antisymmetric under any permutation of identical quarks, the $qq$-cluster spatial-spin-flavor configurations for $qq\bar{q}\bar{q}$ states are $\psi^{osf}_{[11]}$ for the color configuration $\psi^c_{[2]}$ and $\psi^{osf}_{[2]}$ for $\psi^c_{[11]}$. All possible configurations are as follows:
\begin{flalign}\label{conf}
\psi^c_{[2]}\psi^o_{[2]}\psi^s_{[11]}\psi^f_{[2]},\ \psi^c_{[2]}\psi^o_{[2]}\psi^s_{[2]}\psi^f_{[11]},\nonumber \\
\psi^c_{[2]}\psi^o_{[11]}\psi^s_{[11]}\psi^f_{[11]},\
\psi^c_{[2]}\psi^o_{[11]}\psi^s_{[2]}\psi^f_{[2]}
\end{flalign} 
for color sextet $\psi^c_{[2]}$, and 
\begin{flalign}\label{conf}
\psi^c_{[11]}\psi^o_{[2]}\psi^s_{[11]}\psi^f_{[11]} ,\ \psi^c_{[11]}\psi^o_{[2]}\psi^s_{[2]}\psi^f_{[2]}, \nonumber \\
\psi^c_{[11]}\psi^o_{[11]}\psi^s_{[11]}\psi^f_{[2]} ,\ \psi^c_{[11]}\psi^o_{[11]}\psi^s_{[2]}\psi^f_{[11]}
\end{flalign}
for color triplet $\psi^c_{[11]}$, where, $\psi^o_{[2]}$ and $\psi^o_{[11]}$ correspond to $l_{\chi_1}=0$ and $l_{\chi_1}=1$, respectively, as defined by the basis in Table~\ref{spatialbasis}. $\psi^s_{[2]}$ and $\psi^s_{[11]}$ correspond to spin $s_{qq}=1$ and $s_{qq}=0$, respectively. $\psi^f_{[2]}$ and $\psi^f_{[11]}$ correspond to isospin $I_{qq}=1$ and $I_{qq}=0$, respectively; analogous definitions apply to the antidiquark cluster. Thus, the configurations of total isospin $I=0$ and $I=1$ light tetraquarks are listed in Table~\ref{tab:configs_qqqq}.

\begin{table}[tb]
\begin{ruledtabular}
\centering
\small
\caption{$I=0$ and $I=1$ $1^{-+}$ Light tetraquark configurations in a complete harmonic oscillator (HO) basis. Internal orbitals $l_{\chi_1},l_{\chi_2},l_{\chi_3}\in\{0,1\}$; $l_{\chi_1}+l_{\chi_2}+l_{\chi_3}=1$. Pairwise Pauli is enforced for $(12)$ and $(34)$.}
\label{tab:configs_qqqq}
\begin{tabular}{cccccccc}
{Bases} &
\multicolumn{3}{c}{$(qq)$} &
\multicolumn{3}{c}{$(\bar q\bar q)$} &
\multirow{2}{*}{Total $I$} \\
 \{$l_{\chi_1},l_{\chi_2},l_{\chi_3}$\} & $\psi^c_{qq}$ & $s_{qq}$ & $I_{qq}$ & $\psi^c_{\bar q\bar q}$ & $s_{\bar q\bar q}$ & $I_{\bar q\bar q}$  & \\
\noalign{\vskip 5pt}
 \{1,\,0,\,0\} & $\bar{\mathbf 3}_c$ & 1 & 0 & $\mathbf 3_c$ & 0 & 0  & $0$ \\
 \{0,\,1,\,0\} & $\mathbf 6_c$ & 1 & 0 & $\bar{\mathbf 6}_c$ & 0 & 0  & $0$ \\
 \{0,\,0,\,1\} & $\mathbf 6_c$ & 1 & 0 & $\bar{\mathbf 6}_c$ & 1 & 0  & $0$ \\
\noalign{\vskip 5pt}
 \{0,\,0,\,1\} & $\bar{\mathbf 3}_c$ & 1 & 1 & $\mathbf 3_c$ & 0 & 0 & $1$ \\
 \{0,\,1,\,0\} & $\bar{\mathbf 3}_c$ & 1 & 1 & $\mathbf 3_c$ & 1 & 0 & $1$ \\
 \{1,\,0,\,0\} & $\bar{\mathbf 3}_c$ & 1 & 0 & $\mathbf 3_c$ & 1 & 1 & $1$ \\
 \{0,\,0,\,1\} & $\mathbf 6_c$ & 1 & 0 & $\bar{\mathbf 6}_c$ & 0 & 1 & $1$ \\
 \{0,\,1,\,0\} & $\mathbf 6_c$ & 1 & 0 & $\bar{\mathbf 6}_c$ & 1 & 1 & $1$ \\
 \{1,\,0,\,0\} & $\mathbf 6_c$ & 1 & 1 & $\bar{\mathbf 6}_c$ & 1 & 0 & $1$ \\
\end{tabular}
\end{ruledtabular}
\end{table}
For $qq$ pair, color $\bar{\mathbf 3}_c$ is antisymmetric (A), whereas $\mathbf 6_c$ is symmetric (S). Analogous definitions for $\bar q\bar q$ pair, $\mathbf 3_c$ and $\bar{\mathbf 6}_c$ are A and S, respectively. Spin $s=0$ is A and $s=1$ is S. The internal spatial wave function is S for even $l_{\chi}$ and A for odd $l_{\chi}$. In $SU_f(2)$ flavor symmetry, $I=0$ and $I=1$ correspond to flavor A and S configurations, respectively. 
For the configuration in which both $I_{qq}=1$ and $I_{\bar q\bar q}=1$, the total isospin $I$ can take the values $0$, $1$, or $2$, which may lead to ambiguous isospin eigenstates.
This issue is not considered in this work.

For $cc\bar c\bar c$ tetraquarks, the $cc$ cluster has a symmetric flavor component $\psi^f_{[2]}$, so the possible configurations are as follows:
\begin{flalign}\label{conf}
\psi^c_{[2]}\psi^o_{[2]}\psi^s_{[11]}\psi^f_{[2]},\
\psi^c_{[2]}\psi^o_{[11]}\psi^s_{[2]}\psi^f_{[2]}
\end{flalign} 
for color sextet $\psi^c_{[2]}$, 
\begin{flalign}\label{conf}
\psi^c_{[11]}\psi^o_{[2]}\psi^s_{[2]}\psi^f_{[2]},\
\psi^c_{[11]}\psi^o_{[11]}\psi^s_{[11]}\psi^f_{[2]} 
\end{flalign}
for color triplet $\psi^c_{[11]}$, and similarly for charm anti-diquark cluster. The possible configurations of fully charm tetraquarks are listed in Table~\ref{tab:configs_cccc}.

\begin{table}[b]
\begin{ruledtabular}
\centering
\small
\caption{Possible $1^{-+}$ fully charm tetraquark configurations for $1^{-+}$ in a complete HO basis. Internal orbitals $l_{\chi_1},l_{\chi_2},l_{\chi_3}\in\{0,1\}$; $l_{\chi_1}+l_{\chi_2}+l_{\chi_3}=1$.}
\label{tab:configs_cccc}
\begin{tabular}{ccccccccccc}
{Bases} & \multicolumn{2}{c}{$(cc)$} & \multicolumn{2}{c}{$(\bar c\bar c)$}  \\
 \{$l_{\chi_1},l_{\chi_2},l_{\chi_3}$\} & $\psi^c_{cc}$ & $s_{cc}$ & $\psi^c_{\bar c\bar c}$ & $s_{\bar c\bar c}$  \\
\noalign{\vskip 5pt}
\{1,\,0,\,0\} & $\mathbf 6_c$ & 1 &  $\bar{\mathbf 6}_c$ & 0   \\
\{0,\,1,\,0\} & $\bar{\mathbf 3}_c$ & 1 &  $\mathbf 3_c$ & 0   \\
\{0,\,0,\,1\} & $\bar{\mathbf 3}_c$ & 1 &  $\mathbf 3_c$ & 1   \\
\end{tabular}
\end{ruledtabular}
\end{table}

For $qc\bar q\bar c$ tetraquarks, there is no Pauli-antisymmetry constraint tying flavor to color, spin, or space, so no pair contains two identical quarks or two identical antiquarks. The color-spin-orbit (CSO) sector can therefore mix freely, independent of flavor, and charmoniumlike tetraquark states are isospin-blind. In this work, we use the notation $\psi^c_{\bar 3\otimes3}\psi^{s=1}_{(1\otimes1)}$, $\psi^c_{\bar 3\otimes3}\psi^{s=1}_{(1\otimes0)}$, $\psi^c_{6\otimes \bar 6}\psi^{s=1}_{(1\otimes1)}$, $\psi^c_{6\otimes \bar 6}\psi^{s=1}_{(1\otimes0)}$ to denote these configurations. The possible configurations of charmoniumlike tetraquark are listed in Table~\ref{tab:configs_qcqc}

\begin{table}[hptb]
\begin{ruledtabular}
\centering
\small
\caption{Possible $1^{-+}$ charmoniumlike tetraquark configurations in a complete HO basis. Internal orbitals $l_{\chi_1},l_{\chi_2},l_{\chi_3}\in\{0,1\}$; $l_{\chi_1}+l_{\chi_2}+l_{\chi_3}=1$.}
\label{tab:configs_qcqc}
\begin{tabular}{ccccccccccc}
{Bases} &
\multicolumn{2}{c}{$(qc)$} &
\multicolumn{2}{c}{$(\bar q\bar c)$} \\
 \{$l_{\chi_1},l_{\chi_2},l_{\chi_3}$\}  & $\psi^c_{qc}$ & $s_{qc}$  & $\psi^c_{\bar q\bar c}$ & $s_{\bar q\bar c}$  \\
\noalign{\vskip 5pt}
\{0,\,0,\,1\}  & $\bar{\mathbf 3}_c$ & 1 & $\mathbf 3_c$ & 0  \\
\{0,\,1,\,0\} & $\bar{\mathbf 3}_c$ & 1 &  $\mathbf 3_c$ & 0   \\
\{1,\,0,\,0\} & $\bar{\mathbf 3}_c$ & 1 &  $\mathbf 3_c$ & 0   \\
\noalign{\vskip 5pt}
\{0,\,0,\,1\} & $\bar{\mathbf 3}_c$ & 1 &  $\mathbf 3_c$ & 1  \\
\{0,\,1,\,0\} & $\bar{\mathbf 3}_c$ & 1 &  $\mathbf 3_c$ & 1   \\
\{1,\,0,\,0\} & $\bar{\mathbf 3}_c$ & 1 &  $\mathbf 3_c$ & 1   \\
\noalign{\vskip 5pt}
\{0,\,0,\,1\}  & $\mathbf 6_c$ & 1 &  $\bar{\mathbf 6}_c$ & 0   \\
\{0,\,1,\,0\} & $\mathbf 6_c$ & 1 &  $\bar{\mathbf 6}_c$ & 0   \\
\{1,\,0,\,0\} & $\mathbf 6_c$ & 1 &  $\bar{\mathbf 6}_c$ & 0   \\
\noalign{\vskip 5pt}
\{0,\,0,\,1\}  & $\mathbf 6_c$ & 1 &  $\bar{\mathbf 6}_c$ & 1   \\
\{0,\,1,\,0\} & $\mathbf 6_c$ & 1 &  $\bar{\mathbf 6}_c$ & 1   \\
\{1,\,0,\,0\} & $\mathbf 6_c$ & 1 &  $\bar{\mathbf 6}_c$ & 1   \\
\end{tabular}
\end{ruledtabular}
\end{table}

\subsection{Hamiltonian}

We employ a widely used nonrelativistic Hamiltonian~\cite{Wang:2021kfv,Wu:2024ocq,Zhao:2025kno} to study meson and tetraquark systems, which takes the form:
\begin{flalign}\label{eqn::ham}
H= &H_0+ H_{so}, \nonumber \\
H_{0} = &\sum_{k=1}^{N} (\frac12M^{ave}_{k}+\frac{p_k^2}{2m_{k}}) \nonumber \\
&+\sum_{i<j}^{N}(-\frac{3}{16}\vec\lambda_{i}\cdot\vec\lambda_{j})(V_0(r_{ij})+V_{ss}(r_{ij})), \nonumber \\
H_{so} = &\sum_{i<j}^{N}(-\frac{3}{16}\vec\lambda_{i}\cdot\vec\lambda_{j})(V_{so}(r_{ij})),
\end{flalign}
where $M^{ave}_k$ denotes the spin-averaged mass as $\frac{1}{4}M_{PS}+\frac{3}{4}M_V$. $m_k$ are the constituent quark masses. $\vec \lambda_{i}$ are the quark color operator in SU(3). The central potential $V_{0}(r_{ij})$, the spin-spin interaction $V_{ss}(r_{ij})$ and spin-orbit interaction $V_{so}(r_{ij})$ take the form:
\begin{flalign}\label{Hsd}
V_{0}&=A_{ij} r_{ij}-\frac{B_{ij}}{r_{ij}}, \nonumber \\ 
V_{ss}&=\frac{1}{6m_im_j}\Delta V_V(r)\vec\sigma_{i}\cdot\vec\sigma_{j},  \nonumber \\
V_{so} &= \frac{1}{r_{ij}} \frac{dV_{V}}{d(r_{ij})} \frac{1}{4} \Bigg[ \left( \frac{1}{m_i^2} + \frac{1}{m_j^2} + \frac{4}{m_i m_j} \right) \vec{L_{ij}} \cdot \vec{S_{ij}} \nonumber \\
& \quad +\left( \frac{1}{m_i^2} - \frac{1}{m_j^2} \right) \vec{L_{ij}} \cdot \left( \vec{s_i} - \vec{s_j} \right) \Bigg] \nonumber \\
& \quad - \frac{1}{r_{ij}} \frac{dV_{S}}{d(r_{ij})} \left( \frac{\vec{L_{ij}} \cdot  \vec{s_i}}{2m_i^2} + \frac{\vec{L_{ij}} \cdot  \vec{s_j}}{2m_j^2} \right), 
\end{flalign}
where $\vec\sigma_i$ are the quark spin operator in SU(2). Note that we have employed $V_V(r)=-BErf[\sigma r]/r$ and $V_S(r)=Ar$, taken from Ref.~\cite{Schoberl:1986bv}. $m_i$ and $m_j$ are constituent quark masses of ith and jth quark.
$\vec{s_i}$ represents the spin operator for ith quark. $\vec{S_{ij}}=\vec{s_i}+\vec{s_j}$ is the spin operator for (ij)th quark pair. $\vec{L_{ij}}$ is relative orbital operator, taking the form, $\vec{L_{ij}}=\vec{r_{ij}}\times\vec{p_{ij}}$, where $\vec{p_{ij}}=\frac{m_i\vec{p_i}-m_j\vec{p_j}}{m_i+m_j}$.

In line with the previous works~\cite{Zhao:2020jvl,Zhao:2021jss,Zhao:2025kno}, $A_{ij}$, $B_{ij}$, and $\sigma_{ij}$ are taken to be mass-dependent coupling parameters, taking the form
\begin{eqnarray}
A_{ij}= a+bm_{ij}, \;\; B_{ij}=B_0 \sqrt{\frac{1}{m_{ij}}}, \;\; \sigma_{ij} =\sigma_0{m_{ij}}.
\end{eqnarray}
with $m_{ij}$ being the reduced mass of $i$th and $j$th quarks, defined as $m_{ij}=\frac{2 m_i m_j}{m_i+m_j}$. $a$, $b$, $B_0$, and $\sigma_{0}$ are constants.

The central potential $V_0$ and the spin-spin interaction $V_{ss}$ are treated as leading effects, while the spin-orbit interaction $V_{so}$ is regarded as a perturbation that shifts the spectrum. The Schr\"{o}dinger equation is solved for mesons and tetraquarks with the Hamiltonian $H_0$, yielding the eigenvalue $E_0$ and eigenstates $\psi_0$. The mass spectrum is then obtained by adding the diagonalized $H_{so}$ in the basis of these eigenstates.

The coupling constants $a$, $b$, $B_0$, and $\sigma_0$, as well as the constituent quark masses $m_u$ and $m_c$, are taken from previous work~\cite{Zhao:2020jvl,Zhao:2025kno}. These parameters were fixed by reproducing the S- and P-wave mass spectra of light, charmed, bottom, charmonium, and bottomonium mesons within the Hamiltonian in Eq.~(\ref{eqn::ham}). The adopted parameters and constituent quark masses are listed in Table~\ref{tab:coupling}, and the corresponding fits can be found in Ref.~\cite{Zhao:2025kno}.

\begin{table}[hptb]
\caption{\label{tab:coupling}
Imported model coupling parameters.}
\begin{ruledtabular}
\begin{tabular}{ccccccccc}
$a$ & $b$ & $B_0$ & $\sigma_0$ & $m_u$ & $m_c$ \\
\noalign{\vskip 2pt}\hline\noalign{\vskip 2pt}
67413 $\mathrm{MeV}^2$ & 35$~\mathrm{MeV}$ & 31.7~$\mathrm{MeV}^{1/2}$ & 0.7 & 420$~\mathrm{MeV}$ & 1270~$\mathrm{MeV}$ 
\end{tabular}
\end{ruledtabular}
\end{table}

\subsection{Two-body strong decay}

\begin{figure}[hptb]
\centering
\includegraphics[width=0.4\textwidth]{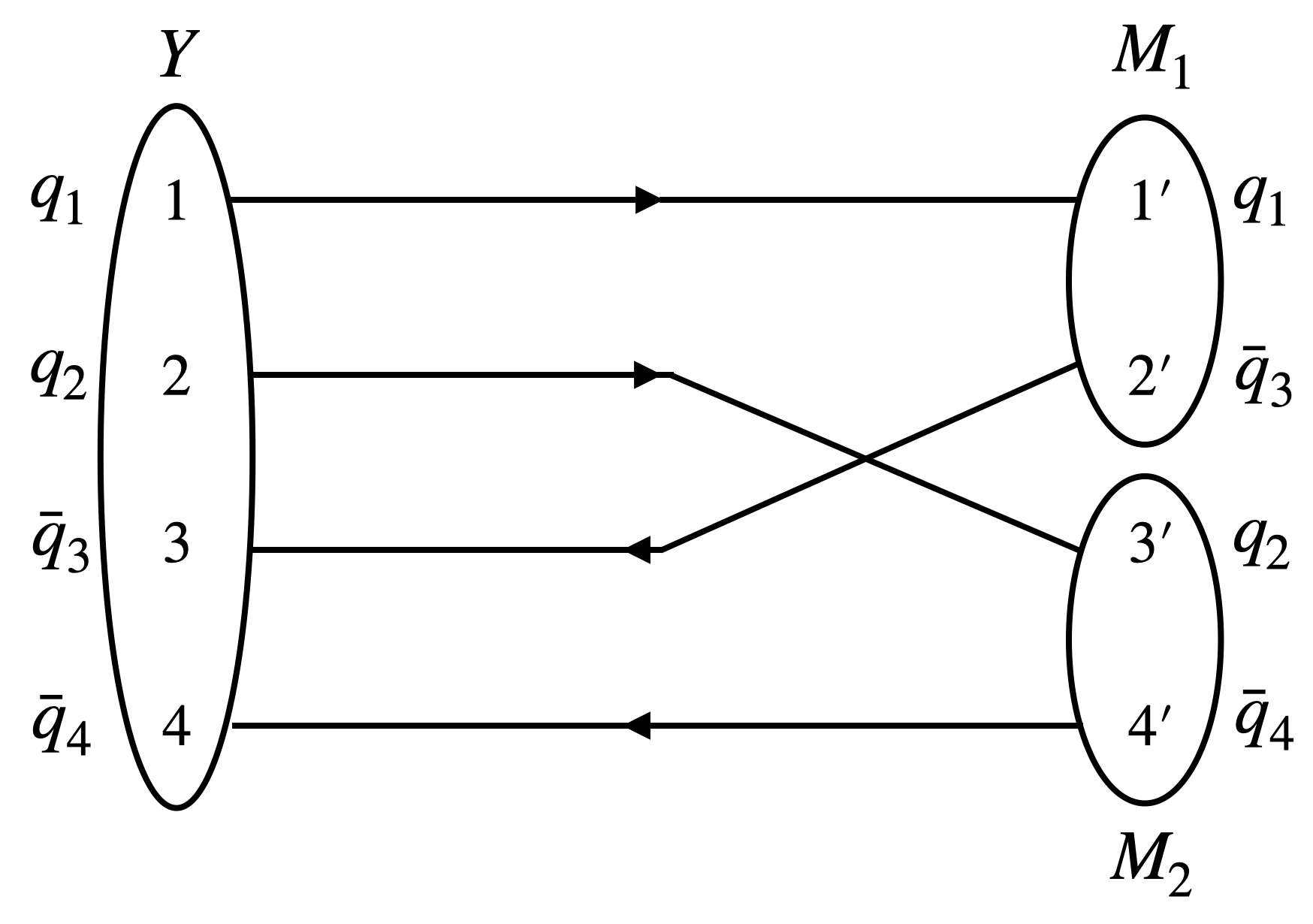}
\caption{Quark rearranged diagram for $1^{-+}$ tetraquark.}
\label{fig:QRD}
\end{figure}

We study the two-body strong decay properties of $1^{-+}$ Y tetraquark states within the rearrangement mechanism, as shown in Fig.~\ref{fig:QRD}, which is believed to be the dominant mechanism for tetraquarks decaying into two mesons. The transition matrix of the color-spin-flavor takes the forms,
\begin{eqnarray}
T_{csf}&=&\langle \psi_f^{csf} | \psi_i^{csf} \rangle \nonumber \\
&=&\langle \psi_{M_1M_2}^{c} | \psi_Y^{c} \rangle \langle \psi_{M_1M_2}^{s} | \psi_Y^{s} \rangle \langle \psi_{M_1M_2}^{f} | \psi_Y^{f} \rangle
\end{eqnarray}
where $\psi_Y^{c}$, $\psi_Y^{s}$, and $\psi_Y^{f}$ are color, spin, and flavor wave functions of the initial tetraquark states taken from the mass spectrum calculation. $\psi^c_{M_1M_2}$, $\psi^s_{M_1M_2}$ and $\psi^f_{M_1M_2}$ are the color, spin, and flavor wave functions of the final states of possible two-body strong decay channels: $\omega h_1$ and $\eta f_1$ for isospin $I=0$ light tetraquarks; $\rho h_1$ and $\pi f_1$ for isospin $I=1$ light tetraquarks; $\pi/\eta+\chi_{c1}$ and $\rho/\omega + h_c$ for charmoniumlike tetraquarks; and $\eta_c \chi_{c1}$ and $J/\psi h_c$ for fully charm tetraquarks.

To partially account for kinematic effects beyond the color-spin-flavor overlap, we further weight each channel by the corresponding two-body phase space evaluated at the tetraquark mass, and estimate the relative partial widths as
\begin{eqnarray}
&\Gamma_\alpha \propto \Phi_\alpha(M_Y)\,|T_{csf}^{(\alpha)}|^2,\nonumber\\
&\Phi_\alpha(M_Y)=k_\alpha\,\frac{E_1(k_\alpha)\,E_2(k_\alpha)}{M_Y},
\end{eqnarray}
where $\alpha$ labels a given decay mode $Y\to M_1M_2$, $M_Y$ is the mass of the initial tetraquark, and $E_{1,2}(k)=\sqrt{m_{1,2}^2+k^2}$ are the on-shell energies of the final mesons with masses $m_{1,2}$. The breakup momentum $k_\alpha$ in the $Y$ rest frame is
\begin{equation}
k_\alpha=\frac{\lambda^{1/2}(M_Y^2,m_1^2,m_2^2)}{2M_Y},\qquad 
\end{equation}
where $\lambda(x,y,z)=x^2+y^2+z^2-2xy-2xz-2yz$.

In the work, the ratios between $\Gamma_\alpha$ of various decay channels are approximated as the decay branching ratios (relative decay ratios), supposing that the contributions of the spatial part of different decay channels are not significantly different. 

\section{\label{sec:RND}Results and Discussion}

\begin{table*}[tb]
\caption{Expectation values of hyperfine interaction of S=1 $qc\bar q\bar c$ $\langle\psi^{k1}_c\psi^{h1}_s|\vec \lambda_i\cdot\vec \lambda_j\vec \sigma_i\cdot\vec \sigma_j|\psi^{k2}_c\psi^{h2}_s\rangle$}
\label{tab:mixture-cs}
\resizebox{\textwidth}{!}{%
\begin{tabular}{lcccc}
\hline
\hline
$\vec \lambda_i\cdot\vec \lambda_j\vec \sigma_i\cdot\vec \sigma_j$ &  $|\psi^c_{\bar 3\otimes 3}\psi^{S=1}_{(1\otimes0)}\rangle$ &  $|\psi^c_{\bar 3\otimes 3}\psi^{S=1}_{(1\otimes1)}\rangle$ & $|\psi^c_{6\otimes \bar 6}\psi^{S=1}_{(1\otimes0)}\rangle$ & $|\psi^c_{6\otimes \bar 6}\psi^{S=1}_{(1\otimes1)}\rangle$ \\
\noalign{\vskip 3pt}\hline\noalign{\vskip 3pt}
$|\psi^c_{\bar 3\otimes 3}\psi^{S=1}_{(1\otimes0)}\rangle$ & $\left(-\frac{8}{3},0,0,0,0,8\right)$&$\left(0,-\frac{4\sqrt{2}}{3},-\frac{4\sqrt{2}}{3},\frac{4\sqrt{2}}{3},\frac{4\sqrt{2}}{3},0\right)$&$\left(0,0,0,0,0,0\right)$&$\left(0,-4,4,-4,4,0\right)$\\
$|\psi^c_{\bar 3\otimes 3}\psi^{S=1}_{(1\otimes1)}\rangle$ &$\left(0,-\frac{4\sqrt{2}}{3},-\frac{4\sqrt{2}}{3},\frac{4\sqrt{2}}{3},\frac{4\sqrt{2}}{3},0\right)$&$\left(-\frac{8}{3},\frac{4}{3},\frac{4}{3},\frac{4}{3},\frac{4}{3},-\frac{8}{3}\right)$&$\left(0,-4,4,-4,4,0\right)$&$\left(0,2\sqrt{2},-2\sqrt{2},-2\sqrt{2},2\sqrt{2},0\right)$\\
$|\psi^c_{6\otimes \bar 6}\psi^{S=1}_{(1\otimes0)}\rangle$ &$\left(0,0,0,0,0,0\right)$&$\left(0,-4,4,-4,4,0\right)$& $\left(-\frac{4}{3},0,0,0,0,-4\right)$ &$\left(0,-\frac{10\sqrt{2}}{3},-\frac{10\sqrt{2}}{3},\frac{10\sqrt{2}}{3},\frac{10\sqrt{2}}{3},0\right)$\\
$|\psi^c_{6\otimes \bar 6}\psi^{S=1}_{(1\otimes1)}\rangle$ &$\left(0,-4,4,-4,4,0\right)$&$\left(0,2\sqrt{2},-2\sqrt{2},-2\sqrt{2},2\sqrt{2},0\right)$&$\left(0,-\frac{10\sqrt{2}}{3},-\frac{10\sqrt{2}}{3},\frac{10\sqrt{2}}{3},\frac{10\sqrt{2}}{3},0\right)$&$\left(\frac{4}{3},\frac{10}{3},\frac{10}{3},\frac{10}{3},\frac{10}{3},\frac{4}{3}\right)$\\
\hline
\hline
\end{tabular}%
}
\end{table*}

The color operator $\vec \lambda_i\cdot\vec \lambda_j$ in the Hamiltonian in Eq.~(\ref{eqn::ham}), along with the mass dependent coefficient in the Cornell-like potential, mixes different color configurations, leading to nonzero cross terms for tetraquarks.

Similarly, the color-spin operator $\vec \lambda_i\cdot\vec \lambda_j\vec \sigma_i\cdot\vec \sigma_j$, together with the mass dependent coefficient in the spin-spin interaction, 
\begin{flalign}\label{AVss}
\vec\lambda_{i}\cdot\vec\lambda_{j}V_{ss}&=\frac{2B_{ij}\sigma_{ij}^3}{3m_im_j\sqrt \pi}e^{-\sigma_{ij}^2r_{ij}^2}\vec\lambda_{i}\cdot\vec\lambda_{j}\vec\sigma_{i}\cdot\vec\sigma_{j}, 
\end{flalign}
mixes different color-spin configurations.

For reference, the expectation values of $O_{ij}=\vec \lambda_i\cdot\vec \lambda_j\vec \sigma_i\cdot\vec \sigma_j$ for $S=1$ states are listed in Table~\ref{tab:mixture-cs}, with all components given in the order $(O_{12},O_{13},O_{23},O_{14},O_{24},O_{34})$. The mass dependent factor in Eq.~(\ref{AVss}) induces nonzero cross terms between different color-spin configurations.

In conclusion, the $I=0$ light tetraquarks are linear combinations of $|\psi^c_{\bar3\otimes3}\psi^{S=1}_{(1\otimes0)},$ 
$\psi^c_{6\otimes\bar6}\psi^{S=1}_{(1\otimes0)},$
$\psi^c_{6\otimes\bar6}\psi^{S=1}_{(1\otimes1)}\rangle$, and $I=1$ light tetraquarks are linear combinations of $|\psi^c_{6\otimes\bar6}\psi^{S=1}_{(1\otimes0)},$
$\psi^c_{\bar3\otimes3}\psi^{S=1}_{(1\otimes0)},$
$\psi^c_{6\otimes\bar6}\psi^{S=1}_{(1\otimes1)},$
$\psi^c_{\bar3\otimes3}\psi^{S=1}_{(1\otimes1)}\rangle$. For fully charm tetraquark, $|\psi^c_{6\otimes\bar6}\psi^{S=1}_{(1\otimes0)},$
$\psi^c_{\bar3\otimes3}\psi^{S=1}_{(1\otimes0)},$
$\psi^c_{\bar3\otimes3}\psi^{S=1}_{(1\otimes1)}\rangle$ configurations are considered due to the quark configurations we have mentioned in Sec.~\ref{sec:TMA}. The charmoniumlike tetraquarks are linear combinations of $|\psi^c_{6\otimes\bar6}\psi^{S=1}_{(1\otimes0)},$
$\psi^c_{\bar3\otimes3}\psi^{S=1}_{(1\otimes0)},$
$\psi^c_{6\otimes\bar6}\psi^{S=1}_{(1\otimes1)},$
$\psi^c_{\bar3\otimes3}\psi^{S=1}_{(1\otimes1)}\rangle$.

We tested convergence with increasing bases and adopt $N_{\max}=13$ (corresponding to $n_{\max}=6$ for $L=1$), which yields stable masses in all three tetraquark configurations.
The masses obtained with this setting are taken as the final results in this section.

\subsection{Results of light tetraquark}

The mass results for the mixed light tetraquark states are given in Table~\ref{tab:lightresult}, where En (n = 1,2,3,...) represents the $n$th eigenstate of the Hamiltonian in Eq.~(\ref{eqn::ham}) with color-spin configuration mixing. The relative decay ratios of the E1 to E5 with various mixed configurations are also listed in Table~\ref{tab:lightresult}, which are normalized to the E3 $\omega h_1$ channel $= 1$ for $I=0$ states and E3 $\rho h_1$ channel $= 1$ for $I=1$ states. THR represents the threshold of two meson final states $M_1+M_2$, with the units of MeV. The symbol ``...'' in table indicates that the predicted mass lies below the corresponding two-body threshold; therefore this decay channel is kinematically forbidden in our framework. 
\begin{table}[hptb]
\caption{\label{tab:lightresult}
Masses of mixed $qq\bar q\bar q$ tetraquark states. The mass results are given in units of MeV. THR denotes the threshold of the two-meson final state $M_1+M_2$, also in units of MeV. The decay ratios are normalized to the E3 $\omega h_1$ channel ($=1$) for $I=0$ states and to the E3 $\rho h_1$ channel ($=1$) for $I=1$ states.}
\begin{ruledtabular}
\begin{tabular}{ccccccccc}
$qq\bar q\bar q$ & THR & E1 & E2 & E3 & E4 & E5  \\
\noalign{\vskip 3pt}\hline\noalign{\vskip 3pt}
$M_{cal}$ $I=0$ &         & 1961 & 2231 & 2242 & 2320 & 2516 \\
$\omega h_1(1170)$ & 1952 & 0.01 & 0.01 & 1 & 32 & 26\\
$\eta f_1(1285)$   & 1835 & 0.03 & 0.01 & 0.1 & 7 & 4 \\
\noalign{\vskip 3pt}\hline\noalign{\vskip 3pt}
$M_{cal}$ $I=1$ &  & 1884 & 1924 & 2112 & 2247 & 2270 \\
$\rho h_1(1170)$ & 1940 & ... & ... & 1 &  1.6  &  2.0 \\
$\pi f_1(1285)$ & 1420 & 38 & 42 & 2.0  &  1.2 &  0.2\\
\end{tabular}
\end{ruledtabular}
\end{table}

{For an S-wave decay of an exotic $1^{-+}$ state, the only two-meson combinations allowed by the quantum numbers are $\left(0^{-+}\oplus1^{++}\right)$ and $\left(1^{--}\oplus1^{+-}\right)$. All other $S+P$ (S-wave + P-wave meson) combinations either fail to satisfy $C=+$ or cannot produce $J=1$ when the relative orbital angular momentum between the two mesons is $L_{rel}=0$.}

{It is worth mentioning that, in principle, the $I=0$ $1^{-+}$ tetraquark can decay through the $\eta\eta'$ channel when the relative orbital angular momentum is $L_{rel}=1$, which allows $J=1$ and negative parity $P=-$.
For a two-meson final state composed of two different self-conjugate mesons $M_1\neq M_2$ (each with a well-defined intrinsic $C_{M_1},\,C_{M_2}=\pm1$), the total two-body C-parity is simply $C_{tot}=C_{M_1}C_{M_2}$, because there is no interchange symmetry between the two distinct mesons. Since $\eta$ and $\eta'$ are distinct, no exchange symmetry factor applies. Thus, for the $\eta\eta'$ channel, the charge conjugation can take $C_{tot}=C_{M_1}C_{M_2}=+$~\cite{Shi:2023sdy,Matveev:2025vjo}.}

However, in the work, the spin-color-flavor factor of the transition amplitude is zero in the two-body strong decay to $\eta\eta'$ from $I=0$ $1^{-+}$ light tetraquarks.

\subsection{$\pi_1(2015)$}

E852 observed two $1^{-+}$ states while studying $\pi^- p$ collisions in the decay channel $f_1(1285)\pi$~\cite{E852:2004gpn} around 1.6 and 2.0 GeV respectively. The higher-mass peak is as $M=2001\pm30\pm92~\mathrm{MeV}$ and $\Gamma=333\pm52\pm49~\mathrm{MeV}$, commonly denoted $\pi_1(2015)$ (a.k.a.\ $\pi_1(2000)$). Later, the $\pi_1(2015)$ state is observed again by E852 in the decay channel $b_1(1235)\pi$~\cite{E852:2004rfa}, with mass and decay width $M=2014\pm20\pm16~\mathrm{MeV}$ and $\Gamma=230\pm32\pm73~\mathrm{MeV}$ which are consistent with that observed in the $f_1(1285)\pi$ final state.

VES performed a similar partial-wave analysis, and the $\pi_1(1600)$ is observed in both $f_1(1285)\pi$ and $b_1(1235)\pi$ decay channels~\cite{Amelin:2005ry}. Their intensity of the $1^{-+}$ partial wave in both channels above 1.9 GeV is considered to be consistent with that observed by E852~\cite{Meyer:2015eta}. However, VES did not comment on this and did not claim the existence of the $\pi_1(2015)$. 

COMPASS recently presented preliminary results of partial-wave decomposition of the diffractive reaction $\pi^- p\to\pi^-\pi^+\pi^-\eta p$~\cite{Spulbeck:2025ath}, and no clear evidence for an excited state as previously reported by E852~\cite{E852:2004gpn} was found, and they highlighted the model dependence of such interpretations.  

A recent light hybrid nonet calculation in a semirelativistic quark potential model organizes the light $1^{-+}$ $\pi_1(1600)$ states into a hybrid nonet with 1669 MeV correspond to the hybrid configurations $(u\bar{u}-d\bar{d})g/\sqrt{2}$. The following analyses accommodate $\pi_1(2015)$ to be first radial excitation of $\pi_1(1600)$ near $2.1~\text{GeV}$~\cite{Chen:2025pvk}. In this view, $\pi_1(2015)$ may have sizable hybrid content but requires confirmation in $f_1\pi$ and $b_1\pi$ with modern amplitude methods. 

Flux--tube models~\cite{Isgur:1984bm} and lattice-QCD~\cite{Dudek:2010wm} place the lightest $1^{-+}$ near $1.7$--$1.9~\text{GeV}$ and expect dominant $S+P$ decays, consistent with the observed $f_1\pi$ and $b_1\pi$ modes. The slightly higher mass of $\pi_1(2015)$ can be explained by hybrid--tetraquark mixing or by identifying it as an excited partner.

{A QCD sum rule study of the light $1^{-+}$ hybrid mesons gives a conservative mass range of $1.7$--$2.6~\text{GeV}$, which naturally accommodates the $\pi_1(2015)$ as a hybrid but disfavors the $\pi_1(1600)$ as a pure hybrid, indicating possible additional nonhybrid components~\cite{Huang:2014hya}. A related light-cone QCD sum rule analysis finds a large $\pi_1\to b_1\pi$ width increasing across the $1.6$--$2.0~\text{GeV}$ region, but a very small $\rho\pi$ width, consistent with the expectation of dominant $S+P$ decays. This suggests that a predominantly hybrid $\pi_1(2015)$ should couple strongly to $b_1\pi$, while the $\rho\pi$ channel remains suppressed~\cite{Huang:2016upt}.}

The observed $S+P$ decay pattern matches expectations for hybrids, yet is also natural for a compact tetraquark. A QCD sum rule analysis of exotic tetraquark states, using diquark-antidiquark currents with $I^GJ^{PC}=1^-1^{-+}$ and $0^+1^{-+}$~\cite{Chen:2008qw,Chen:2008ne} can accommodate $\pi_1(2015)$, with the $qs\bar q\bar s$ configuration around $2.0~\text{GeV}$, which supports the assignment of $\pi_1(2015)$ to a compact tetraquark that naturally prefers $S+P$ decays such as $b_1\pi$ and $f_1\pi$.

{In this work, the $I=1$ light tetraquark E2 state with a mass of around 2 GeV is comparable to the $\pi_1(2015)$ within the experimental uncertainties. The $\pi_1(2015)$ may be assigned to the E2 tetraquark state based on its large decay ratio to $\pi f_1$. We suggest searching for $I=1$ tetraquark states in $S+P$ channels around 1.9--2.0 GeV, and for $I=0$ tetraquark states in $S+P$ channels around 2.3 GeV and 2.5 GeV, motivated by the large decay ratios predicted for several tetraquark states.}

\subsection{$\eta_1(1855)$}

In 2022, the BESIII Collaboration reported the discovery of a new isoscalar exotic state, $\eta_1(1855)$ with quantum numbers $1^{-+}$, observed in the decay process $J/\psi\to\gamma\eta\eta'$ \cite{BESIII:2022riz}. The measured mass is $(1855\pm9^{+6}_{-1})$ MeV, and the width is $(188\pm18^{+3}_{-8})$ MeV, with a statistical significance greater than $19\sigma$.

BESIII also estimated the product branching fraction as $\mathcal{B}(J/\psi\to\gamma\eta_1(1855))\times\mathcal{B}(\eta_1(1855)\to\eta\eta')=(2.70\pm0.41^{+0.16}_{-0.35})\times10^{-6}$ \cite{BESIII:2022riz}. This state is considered a candidate for a hybrid meson, in agreement with lattice QCD predictions \cite{Dudek:2013yja}.

In the hybrid meson framework, the observation of $\eta_1(1855)$, together with the isovector $\pi_1(1600)$, provides valuable information for understanding the $1^{-+}$ hybrid nonet \cite{Qiu:2022ktc,Chen:2022qpd,Shastry:2022mhk,Wang:2022sib,Chen:2022isv,Swanson:2023zlm,Chen:2023ukh,Shastry:2023ths}. A semirelativistic quark potential model was used to estimate the masses of light hybrid mesons with $1^{-+}$ \cite{Chen:2023ukh}. This study suggests that $\pi_1(1600)$ and $\eta_1(1855)$ might correspond to the hybrid configurations $(u\bar{u}-d\bar{d})g/\sqrt{2}$ and $s\bar{s}g$, respectively.

However, another recent study modeled hybrid mesons as three-body systems (quark-antiquark-gluon), treating the gluon as a real constituent \cite{Ma:2025cew}. Their results indicate that both $\pi_1(1600)$ and $\eta_1(1855)$ cannot be interpreted as $1^{-+}$ hybrids at the same time, due to inconsistencies in their decay widths. In particular, they found that the decay width of $\eta_1(1855)\to\eta\eta'$ is nearly zero at leading order, which challenges its hybrid interpretation.

Alternatively, in the hadronic molecule picture, $\eta_1(1855)$ could be a bound state of $K\bar{K}_1(1400)$, as its mass is about 40 MeV below this threshold \cite{Dong:2022cuw, Yang:2022rck}. Ref. \cite{Dong:2022cuw} used the one-boson exchange model to explore the binding and decay properties of such a molecule and found results supporting this interpretation. Similarly, Ref. \cite{Yang:2022rck} studied its radiative and strong decays using the effective Lagrangian approach and compared them with experimental data.

On the other hand, a study using the chiral quark model did not support the $K\bar{K}_1$ molecular interpretation of $\eta_1(1855)$ \cite{Tan:2025ahx}. In another analysis using QCD sum rules \cite{Liu:2024lph}, the exotic $K\bar{K}_1(1270)$ and $K\bar{K}_1(1400)$ molecular states with $1^{-+}$ were investigated, but the spectral functions turned out negative in physical regions, making the results unreliable.

A tetraquark molecular picture was also considered. In Ref. \cite{Wan:2022xkx}, configurations like $[1_c] s\bar{s}\otimes[1_c] q\bar{q}$ and $[1_c] s\bar{q}\otimes[1_c] s\bar{q}$ were studied using QCD sum rules. The mass calculations suggest that $\eta_1(1855)$ could match the $[1_c] s\bar{s}\otimes[1_c] q\bar{q}$ configuration.

{In this work, the theoretical relative decay ratios for the decay of the $1^{-+}$ light tetraquark state into $\eta\eta'$ are found to be nearly zero. Consequently, the $\eta_1(1855)$ state observed in the $\eta\eta'$ channel~\cite{BESIII:2022riz} cannot be accommodated within the present tetraquark picture when both the mass spectrum and decay properties are taken into account.} 

{The $\eta_1(1855)$ may instead correspond to other configurations, such as a hadronic molecular state or a hybrid meson, as discussed above. In particular, if the $\eta_1(1855)$ is interpreted as a $K_1\bar{K}$ hadronic molecule and decays into $\eta \eta^{\prime}$, the corresponding decay width is expected to be sizable, as discussed in detail in Ref.~\cite{Yan:2023vbh}. While some models support such exotic interpretations, others challenge these assignments, highlighting the need for further theoretical and experimental studies.}

\subsection{Results of charmoniumlike $1^{-+}$ tetraquark}

Lattice QCD predicts that the lightest $1^{-+}$ charmonium hybrid ($\eta_{c1}$) lies around $4.2-4.4~\mathrm{GeV}$ and decays mainly to open-charm $S+P$ channels such as $D_1 \bar D$, $D \bar D$, and $D \bar D^*$. Its hidden-charm branching fractions are small, which makes it hard to observe in clean but suppressed radiative modes~\cite{Shi:2023sdy}.

At present, no charmoniumlike state with $1^{-+}$ has been confirmed. Extensive studies of the $XYZ$ sector by BESIII, LHCb, Belle, and Belle II have found many states with nonexotic, non-$1^{-+}$ quantum numbers (e.g.\ $1^{++},1^{+-},0^{++},2^{++}$), while searches for $1^{-+}$ have so far found nothing. A dedicated BESIII search, $e^+e^-\to\gamma\eta^{(\prime)}\eta_c$ at $\sqrt{s}=4.258\text{--}4.681~\mathrm{GeV}$, saw no significant signal and set 90\% C.L. upper limits on the cross sections~\cite{BESIII:2025etaetac}.

Promising search modes that act as quantum-number filters for $C=+$, odd parity, and $J=1$ are: (i) radiative production at BESIII/Belle~II, $e^+e^-\to\gamma Y(1^{-+})$ with $Y\to S{+}P$ meson pairs ($\eta+\chi_{cJ}$, $\rho+h_c$), which provide clean angular observables for amplitude fits~\cite{Chao:2013cca,Zhang:2025gmm}; and (ii) $B$-meson decays at LHCb/Belle~II, $B\to KY(1^{-+})$ with $Y\to S{+}P$ meson pairs, enabling direct tests of the exotic $1^{-+}$ hypothesis.

Higher-luminosity scans and better Electromagnetic Calorimeter (EMC) calibrations at BESIII (below 4.66~GeV) and Belle~II (10.58~GeV), plus targeted angular-moment analyses for $1^{-+}$ are especially promising~\cite{Yuan:2021wpg}. Notably, BESIII's first direct production of a $C=+$ state, $e^+e^-\to\chi_{c1}$, shows that $e^+e^-$ experiments can access $C=+$ states~\cite{BESIII:2022mtl}.
The work confirms key tools, photon reconstruction and interference control required for a $1^{-+}$ search.

The mass spectrum and S-wave relative decay ratios of the charmoniumlike tetraquarks are listed in Table~\ref{tab:resultqcqc}, and the ratios are normalized to the E3 $\eta \chi_{c1}$ channel $= 1$. Unlike the cases for light tetraquarks and fully charm tetraquarks, where the flavor configurations impose constraints on color-spin-orbit ones, the color-spin-orbit configurations for charmoniumlike tetraquark are independent of flavors. Therefore, the present work is isospin-blind for charmoniumlike tetraquark states. 
As a result, the masses of the $I=0$ and $I=1$ states are the same in this work, and both the $\pi/\eta+\chi_{c1}$ and $\rho/\omega+h_c$ decay channels are included. The relative decay ratios differ due to phase-space factors.

The lowest mass is around 4.2~GeV with a relatively large ratio to 
$\eta \chi_{c1}$ channel. In the work, the theoretical results of relative decay ratios for $1^{-+}$ charmoniumlike tetraquark state decaying to $\eta \eta_c$ are very small. The work suggests searching for tetraquark states in $\eta \chi_{c1}$ channels in the mass range of $4.2-4.3$~GeV due to the large decay ratio of two tetraquark states.

\begin{table}[tb]
\caption{\label{tab:resultqcqc}
Masses of mixed charmoniumlike tetraquark states. The theoretical mass results $M_{cal}$ are given in units of MeV. THR denotes the threshold of the two-meson final state $M_1+M_2$, also in units of MeV. The decay ratios are normalized to the E3 $\eta \chi_{c1}$ channel ($=1$).} 
\begin{ruledtabular}
\begin{tabular}{ccccccccc}
$qc\bar q\bar c$ & THR & E1 & E2 & E3 & E4 & E5 & E6
\\
\noalign{\vskip 3pt}\hline\noalign{\vskip 3pt}
$M_{cal}$ $I=0$ & & 4245 & 4287 & 4289 & 4307 & 4421 & 4441\\
$\eta\chi_{c1}$ & 4057 & 13 & 23 & 1 & 4.5 & 0.1 & 0.6 \\
$\omega h_c$ & 4307 & ... & ... & ... & 0.03 & 0.3 & 0.6 \\
\noalign{\vskip 3pt}\hline\noalign{\vskip 3pt}
$M_{cal}$ $I=1$ & & 4245 & 4287 & 4289 & 4307 & 4421 & 4441\\
$\pi\chi_{c1}$ & 3650 & 27 &  41 &  1.8 &  8 & 0.1 & 0.8 \\
$\rho h_c$ & 4295 & ... & ... & ... & 0.2 & 0.3 & 0.7\\
\end{tabular}
\end{ruledtabular}
\end{table}

\subsection{Results of fully charm $1^{-+}$ tetraquark}

To date there is no confirmed observation of a $1^{-+}$ fully charm state. The established structures in the di-$J/\psi$ spectrum, most prominently the narrow $X(6900)$ and companion peaks near $6.6$ and $7.1$~GeV observed by LHCb~\cite{LHCb:2020bwg}, CMS~\cite{CMS:2023owd}, and ATLAS~\cite{ATLAS:2023bft}, are studied in the previous work~\cite{Zhao:2020jvl}, and tentatively assigned as S-wave tetraquark states.
A recent CMS angular analysis determines the parity and charge-conjugation of the fully charm peaks to be $2^{++}$ and disfavors $J=1$ assignments, thus excluding $1^{-+}$ for those structures~\cite{CMS:2025fpt}. 
Some theoretical studies typically locate low-lying fully charm tetraquarks with $0^{++}$, $2^{++}$ (and sometimes $1^{+-}$), not $1^{-+}$~\cite{Karliner:2020dta,Yang:2021zhe}.

The most sensitive venues are CMS/ATLAS and LHCb, which provide large prompt-charm samples and clean muon triggers for dicharmonium studies. However, the main channel $J/\psi J/\psi$ favors even-$J$ and $C=+$ waves; $J=1$ in S-wave is strongly suppressed near threshold and disfavored by the Bose symmetry of two identical vectors, so sensitivity to $1^{-+}$ is low. Inclusive production also dilutes signals with large nonresonant di-charmonium backgrounds, and limited data in $S+P$ modes hinder full amplitude analyses. A better strategy is to move beyond identical-vector final states to $S+P$ charmonium pairs with definite $C=+$ and richer angular structure (e.g.\ $\eta_c \chi_{cJ}$, $J/\psi h_c$).

On the detector side, LHC Run~3 upgrades, higher-rate di-muon triggers, better mass resolution, and dedicated di-charmonium selections, should increase sensitivity~\cite{CMS:2023owd,ATLAS:2025nsd}. At the $\bar p p$ threshold, PANDA will enable formation scans with excellent resolution and access to exclusive $S+P$ final states, which could be decisive for exotic $J^{PC}$ tests~\cite{Pelizaeus:2012cc}. Belle~II can also contribute via double-charmonium production at 10.58~GeV, with full-amplitude analyses in $J/\psi h_c$ and $\eta_c \chi_{cJ}$, though cross sections are lower than at the LHC.

\begin{table}[tb]
\caption{\label{tab:resultcccc}
Masses of $1^{-+}$ $cc\bar c\bar c$ tetraquark states. The mass results are given in units of MeV. THR denotes the threshold of the two-meson final state $M_1+M_2$, also in units of MeV. The decay ratios are normalized to the E3 $J/\psi h_c$ channel ($=1$).}
\begin{ruledtabular}
\begin{tabular}{ccccccccc}
$cc\bar c\bar c$ & THR & E1 & E2 & E3 & E4 & E5 & E6
\\
\noalign{\vskip 3pt}\hline\noalign{\vskip 3pt}
$M_{cal}$ & & 6595 & 6762 & 6881 & 6908 & 7063 & 7079\\
$J/\psi h_c$ & 6622 &  ... & 0.1 & 1 & 36 & 47 & 0.2 \\
$\eta_c \chi_{c1}$ & 6496 & 2 & 0.1 & 1.5 & 51 & 60 & 0.3 \\
\end{tabular}
\end{ruledtabular}
\end{table}

{The mass spectra and S-wave relative decay ratio are listed in Tables~\ref{tab:resultcccc} for the fully charm tetraquark, and the ratios are normalized to the E3 $J/\psi h_{c}$ channel $= 1$. The lowest mass is around 6.6~GeV with a relatively large ratio to $\eta_c \chi_{c1}$ channel. The work suggests searching for tetraquark states in $\eta_c \chi_{c1}$ channels around 6.6~GeV, and in $J/\psi h_c$ and $\eta_c \chi_{c1}$ channels in the mass range of 6.9-7.1~GeV, due to the large decay ratio of several tetraquark states.}

\section{Summary}\label{sec:SUM}

While the hybrid interpretation has been extensively explored in the literature, the tetraquark scenario remains particularly
compelling due to the rich internal dynamics inherent to four-quark systems.

The masses and decay patterns of $1^{-+}$ light, charmoniumlike, and fully charm tetraquark states are studied using a constituent quark model that incorporates a Cornell-like potential and the Breit-Fermi interaction. The theoretical results for tetraquarks are compared with those of $1^{-+}$ exotic states.

The lowest $1^{-+}$ light tetraquark state is around 1.9 GeV, which is significantly above the observed $\pi_1(1400)$ and $\pi_1(1600)$ states. 

The observed states $\eta_1(1855)$ and $\pi_1(2015)$ are located within the light tetraquark range. 
$1^{-+}$ Y tetraquark states decaying to $\eta \eta'$ is strongly suppressed in our CSF overlap within the rearrangement mechanism.
Thus, the $\eta_1(1855)$ state observed from $\eta\eta'$ by BESIII cannot be accommodated in the current tetraquark picture.

The $\pi_1(2015)$ is assigned as $I=1$ E2 $1^{-+}$ light tetraquark state according to both mass agreement and the large decay ratio.
More experimental data and theoretical studies are essential for making unambiguous assignments.

For the $1^{-+}$ charmoniumlike tetraquark state, the lowest mass is predicted near 4.2 GeV and could decay into $\eta\chi_{c1}$ channel.

The lowest $1^{-+}$ fully charm tetraquark state is around 6.6 GeV and may be searched in $\eta_c \chi_{c1}$ channel.
 
\begin{acknowledgments}

This research has received funding support from the NSRF via the Program Management Unit for Human Resources \& Institutional Development, Research and Innovation [grant number~B50G670107]. A. L. and Y. Y. acknowledge the support of (i) Suranaree University of Technology, (ii) Thailand Science Research and Innovation (TSRI), and (iii) NSRF, Project No. 195242. N.~T. acknowledges his PhD scholarship from the Development and Promotion of Science and Technology Talents project (DPST), Thailand.

\end{acknowledgments}

\bibliography{arXivPwtetra2025}

\end{document}